# 石墨烯在 Ni(111)表面生长机理的紧束缚计算研究

周辰，胡靖，田圆，赵倩莹，缪灵*，江建军

（华中科技大学电子科学与技术系，武汉 430074）

**摘要** 本文采用 SCC-DFTB 方法，研究了石墨烯在 Ni 金属(111)表面上的生长机理及在台阶面生长情况. 结果分析表明，苯环在 Ni 表面吸附时以界面 fcc 构型总能最低，结构最为稳定. 边缘生长时，附着在衬底表面上的石墨烯层中 C 原子活性从边缘向中间逐渐降低. 在由(111)晶面和(1-11)晶面相交形成的台阶面上，石墨烯片层可连续生长，同时相对衬底表面发生一定偏转，在较大面积时将出现缺陷. 改善石墨烯与衬底台阶处的界面不匹配情况将有利于其大面积高质量生长.

**关键词** 石墨烯；金属衬底；表面结构；台阶生长；紧束缚近似

# Tight-binding calculation of growth mechanism of graphene on Ni(111) surface

Zhou Chen, Hu Jing, Tian Yuan, Zhao Qianying，Miao Ling*，Jiang Jianjun

accpeted by Journal of Atomic and Molecular Sciences (in the press, in Chinese)

（Department of Electronic Science and Technology, Huazhong University of Science and Technology, Wuhan 430074, China）

**Abstract:** The nucleation of graphene on Ni surface, as well as on the step, is studied using a tight binding method of SCC-DFTB. The result demonstrates that the fcc configuration has the lowest total energy and thus is the most stable one compared to the other two structures when benzene ring is absorbed on the Ni(111) surface. The activity of marginal growth graphene's carbon atoms decreases from the boundary to the center, when they are absorbed on the substrate. Graphene layer can grow continuously on step surface formed by intersection of Ni(111) and Ni(1-11) surface. Meanwhile, a mismatch will occur between the layer and Ni surface and thus leads to flaws when the layer grows larger. Reducing the mismatch between the graphene and the step surface will benefit the growth of graphene of large area and high quality.

---

# 1 引言

2004 年石墨烯被发现后[1],迅速成为了物理、化学和材料等等学科的研究热点,在很多领域中有极大应用前景[2][3]. 高质量、大尺寸石墨烯材料的可控制备,对开展石墨烯本征物理特性及其应用研究,都具有重要的科学与技术上的价值. 目前制备石墨烯的方法主要有微机械剥离法[4]、氧化还原法[5]、晶体外延生长法[6]、过渡金属催化的化学气相沉积法[7]等等. 其中化学气相沉积法以金属晶体作为衬底,可通过衬底选择、生长温度、前驱物暴露量等生长参数对石墨烯的生长进行较好调控[8].

当石墨烯和金属衬底的晶格匹配较好时,易于在衬底表面生长形成附着力较大、质量较好的大面积石墨烯. 2008 年,Yu Q 等[9]在金属 Ni 衬底上制备了高质量的石墨烯,并可控制石墨烯薄膜的厚度和缺陷. 2009 年 Kim 等人[10]实现了在多晶 Ni 薄膜上外延生长厘米量级石墨烯. 随着这种方法的发展,单层石墨烯或者是双层石墨烯现在已经可以覆盖 87% 的 Ni 表面[11]. Wang B 等人[12]计算研究了石墨烯在金属表面外延生长的特性. Xu Z 等[13] 利用第一性原理计算方法研究了石墨烯吸附在 Cu(111) 和 Ni(111) 界面上的结构稳定性.

探究石墨烯在金属衬底上的生长机理,对今后的石墨烯大规模、高质量的制备具有指导意义. 石墨烯的生长主要是衬底金属催化分解甲烷等气体产生 C 原子,继而不断在衬底表面成核长大的结果. 在活泼金属如 Pd[14]和 Ru[15]衬底表面,甲烷可以直接分解产生 C 原子. 在 Ni 表面,由于 C 在 Ni 中的高溶解度,甲烷分解产生 C 原子也是顺利的[16]. 然而在 Cu 等较惰性金属表面,甲烷直接完全脱氢较为困难,倾向于由 $CH_x$ 结构在衬底表面成核生长[17].

由于衬底在台阶处的表面活性会高于平面处,故石墨烯在台阶处的生长会更容易. Coraux J 等[18]利用低压化学气相沉积方法在 Ir(111) 上面生长了单层的石墨烯时,发现吸附表面有台阶状的结构,同时获得了杂质含量很少的石墨烯. Loginova E 等[19]揭示了在低碳含量情况下,石墨烯的生长主要发生在金属台阶边缘附近,当碳浓度较高时,则倾向于平台和台阶边缘上. 2010 年 Sprinkle M 等[20]研究发现预处理后 SiC 晶体在 $(1\bar{1}0n)$ 台阶面生长出高质量的石墨烯纳米带. Günther S 等[21]通过实验研究 Ru(0001) 表面外延生长石墨烯时,发现金属台阶面处生长出更有序的单晶石墨烯薄膜,密度泛函计算验证了这种单台阶生长模式

是石墨烯同金属间相互作用的结果.

此前关于石墨烯在衬底上制备的实验研究较多，而其生长机理尚未完全明确，斜坡台阶状金属表面生长石墨烯的研究更为少见. 本文利用紧束缚计算方法，通过进行几何优化及吸附能的计算探讨石墨烯和 Ni 衬底的相互作用规律，研究石墨烯在 Ni 衬底上生长时边缘活性及斜坡上的生长情况.

## 2　计算方法

本文计算采用基于自洽的电荷密度泛函紧束缚（SCC-DFTB）方法的 DFTB+ 软件包[22,23]. SCC-DFTB 是一种起源于密度泛函理论的紧束缚方法，它对交换积分进行近似和参数化[24]，紧束缚近似简化后的哈密顿矩阵元由电荷自洽决定[25]. 计算中势函数为 ZnOrg-0-1 和 pbc-0-3[26]. 结构弛豫使用共轭梯度算法优化了所有原子位置，原子受力的收敛精度为 0.05 eV/nm. 电子自洽计算的能量收敛判据为 $2.72 \times 10^{-4}$ eV.

相对于分子动力学方法，紧束缚方法能够考虑电荷间转移，适用范围更广，精度更高，更适合本文中的模型结构. 相对于基于密度泛函理论的第一性原理计算方法[27][28][29]，紧束缚方法计算速度很高，可以计算较大体系，在成熟的参数条件下，定量上比较准确[30][31][32].

## 3　结果与讨论

本文采用苯环作为计算的对象模拟石墨烯，探究一个苯环在 Ni 衬底上的生长情况. 由于石墨烯生长的方式为边缘不断向外生长，故本文采取在边缘增加 C 原子的方法研究其生长特性. 在研究斜坡生长时，我们选取与 Ni（1 1 1）面晶格相似的 Ni（1 1̄ 1）面与 Ni（1 1 1）面构成斜坡.

### 3.1 Ni(111)表面的苯环吸附

石墨烯在衬底表面生长方式为由小的成核单元边缘不断向外生长. 考虑到碳六元环为石墨烯二维结构最小重复单元，选取苯环作为初始成核单元，探究石墨烯在 Ni(111)表面吸附规律. Ni 晶体为一个面心立方结构，其（111）面循环周期为 3 层. 由于衬底外表面对石墨烯片层的影响最为主要，在此建立 4 层 Ni 原子层的衬底结构. 根据界面处表面原子相对位置的不同，石墨烯在 Ni(111)上有三种稳定吸附位置，分别是 hollow、fcc、hcp，苯环中心分别正对着衬底表面第一、二、三层的 Ni 原子[33]. 作为例子，图 1 给出了 fcc 结构模型，苯环平行放置在

Ni 衬底上方. 下面分别针对这三种不同相对位置进行计算以得到最稳定结构.

结构优化时，苯环 C 原子和顶层 Ni 原子可以自由移动，同时固定下面三层 Ni 原子位置不变. 优化前后苯环和 Ni 衬底原子结构几乎没有变化，只是两者间距 d 有所变化. 三种吸附位置的间距 d 均约为 0.30 nm，略小于范德瓦尔斯作用类型的石墨烯层间距 0.34 nm. 这说明苯环这种边缘饱和 C 环结构与衬底表面原子间相互作用较弱.

进一步计算了体系吸附能 $E = E_0 - E_1 - E_2$ 来定量描述吸附强弱. 式中，$E_0$ 为结构总能，$E_1$ 为苯环能量，$E_2$ 为 Ni 衬底能量. 三种吸附位置 hollow、fcc、hcp 的吸附能分别为 −1.0 eV、−1.3 eV、−1.2 eV. 由于体系稳定性与吸附能大小成正比，故三种结构的稳定性大小关系为: fcc > hcp > hollow. 即 fcc 结构在三种结构中的吸附能最高，稳定性最好，即意味着在 Ni（111）衬底表面，石墨烯最可能以这种结构开始生长. 下面以此结构进一步研究石墨烯生长过程.

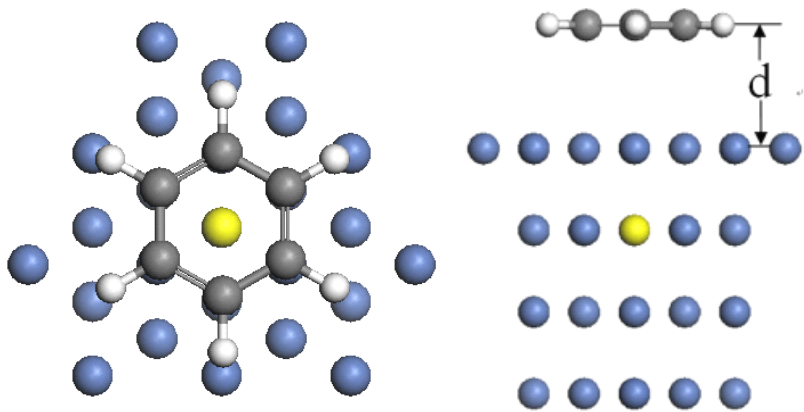

图 1  Ni 衬底吸附苯环的 fcc 结构模型，俯视图（左）与侧视图（右）

Fig.1 fcc structure of benzene ring absorbed on Ni, top view(left) and side view(right)

### 3.2 不同边缘原子活性时的生长

在石墨烯生长过程中，边缘原子活性对其生长扩展起到关键作用. 在此研究不同边缘原子活性时石墨烯的生长情况. 模型为在苯环的三个对称方向上分别添加单个 C 原子，探讨新添加的边缘 C 原子与衬底 Ni 原子的作用规律. 在实际的较大面积石墨烯生长过程中，位于大片石墨烯中心的 C 原子活性与模型中心 6 个 C 原子相同，实际边缘 C 原子活性与模型中边缘的 3 个碳原子相同. 进一步加不同数目 H 原子来分别饱和边缘 C 原子以得到不同边缘原子活性，具体为不加

氢（$C_9H_3$）、加一个氢（$C_9H_6$）、加两个氢（$C_9H_9$）结构，进而探究石墨烯边缘的生长情况.

如图 2 所示，分别结构优化后，三种不同活性模型 $C_9H_n$ 结构的边缘原子成键情况有很大差异. 表 1 给出了对应结构的 C-Ni 键长及吸附能. 可以看出，未加 H 钝化的边缘 C 原子由于饱和程度最低，活性最大，与衬底顶层 Ni 原子作用最强，形成三个 C—Ni 键. 而且由于较强边缘相互作用和较短的 C—Ni 键(0.159 nm)，石墨烯 C 原子和顶层 Ni 原子排布均出现了一定程度的变形. 由前面的计算可知在苯环和第一层 Ni 之间距离为 0.316 nm，几乎是 0.159 nm 的两倍. 所以 $C_9H_3$ 的边缘 C 原子与 Ni 原子成键很强.

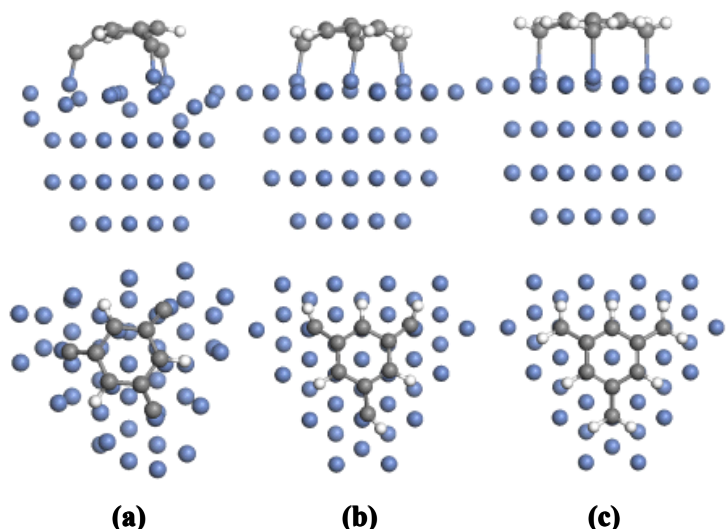

**(a)**      **(b)**      **(c)**

图 2 (a) $C_9H_3$、(b) $C_9H_6$ 和 (c) $C_9H_9$ 优化结构，侧视图（上）和俯视图（下）

Fig. 2 The optimized structures of (a) $C_9H_3$, (b) $C_9H_6$ and (c) $C_9H_9$, side view (up) and top view (down)

表 1   $C_9H_n$ 结构的 C-Ni 键长 $L_{C-Ni}$（/nm）及吸附能 $E_b$（/eV）

Table 1   The C-Ni bond length of $L_{C-Ni}$ (/nm) and adsorption energy $E_b$ (/eV) of the structure of $C_9H_n$

| 结构 | $C_9H_3$ | $C_9H_6$ | $C_9H_9$ |
| --- | --- | --- | --- |
| $L_{C-Ni}$ | 0.159 | 0.182 | 0.218 |
| $E_b$ | 1.45 | 0.78 | 0.64 |

$C_9H_6$ 结构为边缘 C 加一个氢，可视为石墨烯进一步生长的过渡结构，原先最边缘的 C 原子逐渐变为内部 C 原子. 此时边缘 C 原子仍与 Ni 明显成键，$C_9H_6$ 结构中的平均键长为 0.182 nm，表层 Ni 原子的平整度高于 $C_9H_3$，可见边缘 C 原子活性不如 $C_9H_3$. 对于边缘 C 原子加两个氢的结构（$C_9H_9$），这时三个 C 原子的

成键饱和情况就与实际石墨烯内部 C 原子相似. 如图 2 所示，石墨烯层和顶层 Ni 原子明显比 $C_9H_3$ 结构和 $C_9H_6$ 结构平整了许多，与前面 fcc 结构相似度更高.

根据表 1，边缘 C-Ni 键长变化的规律为 $L\text{-}C_9H_3 < L\text{-}C_9H_6 < L\text{-}C_9H_9$. 同时石墨烯中间部分也逐步升高，形成平整石墨烯片层结构. 可知随着石墨烯生长，原先处于边缘的 C 原子逐渐变化为中间 C 原子，其活性也是逐渐降低. 这一点也可以从表面吸附能中体现出来. 吸附能大小关系为 $E_b\text{-}C_9H_3 > E_b\text{-}C_9H_6 > E_b\text{-}C_9H_9$，吸附能越高其边缘 C 原子的活性越大，与键强的规律是一致.

### 3.3 斜坡状表面的石墨烯生长

衬底上斜坡或台阶处金属原子活性一般高于平面处原子，在其上石墨烯成键生长往往更容易. 研究斜坡状表面上石墨烯的生长规律对于制备石墨烯器件及制备特定形状的石墨烯具有重要意义. 基于此，本文进一步研究了斜坡上石墨烯的生长情况.

由于晶体对称性，与 Ni(111)晶面相同的晶面有($\bar{1}$11)、(1$\bar{1}$1)和($\bar{1}\bar{1}$1)面. 为探究石墨烯在衬底台阶上生长情况，即能否在台阶上生长出仍然吸附在衬底上的弯折石墨烯层，构建了结构模型如图 3 所示. 在 Ni(111)面与 Ni(1$\bar{1}$1)面构成的斜坡状表面两平面上分别添加与衬底构成 fcc 结构匹配的石墨烯层. 对于边缘 C 原子，采取加 H 钝化操作.

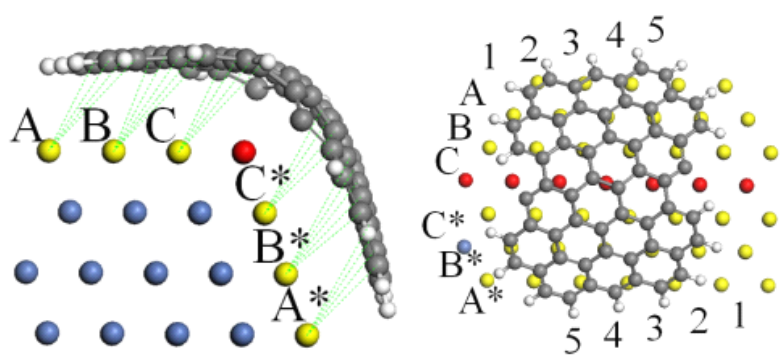

图 3 斜坡上石墨烯的生长情况
Fig. 3 The growth condition of graphene on slope

结构优化后，在两不同晶面上方的分立石墨烯连接生长在了一起，并且石墨烯层仍然平整的吸附在 Ni 衬底上. 然而与单独(111)晶面时 C-Ni 原子匹配很好情况不同，石墨烯层与 Ni 衬底层有一个明显的偏转错位. 如图 3 左图所示，虚线

为 C 原子与相对应的 Ni 原子间的连线,从侧面看过去为伞状散开,这是由于石墨烯层同 Ni 衬底层间存在着旋转偏移,致使连线也随之偏移. 如图 3 右图所示, 红色标示了两平面相交处的 Ni 原子,锯齿状 C 原子链与红色的碳原子链有一个明显的角度差.

图 4 给出了 C 原子与对应 Ni 距离变化关系. 可以看出,在平面方向上,C 原子与对应 Ni 间距离自左向右的变化趋势为逐渐减小,但基本均大于 0.316 nm. 而在斜面上自左向右的变化趋势正好相反,这是因为石墨烯与 Ni 衬底间的相对偏转关于面夹角中分线二次旋转对称. 从数据的值也可以看出平面与斜坡面情况基本对称. 斜面上 C-Ni 间距也基本大于 0.316 nm. 而且在两个晶面上,石墨烯层与 Ni 层的角度偏差均为 11°对称. 说明在两个一定夹角的对称晶面上,石墨烯可以发生同等程度的旋转后连接在一起.

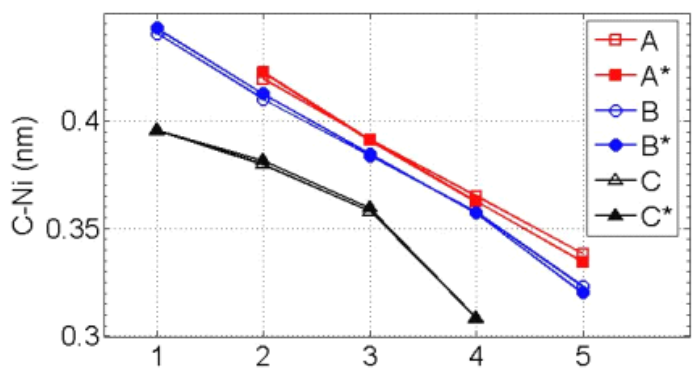

图 4  C 原子与相对应的 Ni 原子的间距

Fig. 4 The distance between C atom and the corresponding Ni atom

同时无论是在平面,还是在斜坡面,随着 C 原子距晶面交汇处的距离越来越远,其与对应 Ni 原子间的间距也越来越大. 这说明,石墨烯与 Ni 衬底表面不匹配程度越来越大. 当石墨烯的面积较大时,其与衬底的不匹配程度将过大,增加了石墨烯缺陷存在的可能性,影响石墨烯的生长稳定性. 增加台阶处石墨烯同衬底金属晶格间的匹配程度,将会是实现石墨烯在台阶上大面积高质量生长的重要条件.

## 4  结论

本文采用 SCC-DFTB 方法研究了石墨烯在 Ni 衬底表面上的生长机理,并进

一步研究了石墨烯在衬底斜坡状台阶处的生长情况. 通过对苯环在 Ni(111)面上三种不同稳定吸附位置 hollow，fcc，hcp 的结构优化，发现 fcc 结构的吸附能最大，结构最稳定. Ni(111)面上的苯环可以吸收边缘 C 原子，从而可以逐步生长为石墨烯，C 原子活性从边缘到中央逐渐降低. 当 Ni 斜坡的平面和斜坡面具有相同的晶格结构时，石墨烯在两个面上表现出可连续生长的特性，但会出现一定程度的偏差错位，可以预测，在大面积斜坡上生长石墨烯片层可能会出现较多的缺陷. 因此，对于金属多晶衬底，提高石墨烯与台阶状金属衬底的匹配程度，将会成为减少缺陷，提高稳定性的重要途径.

## 参考文献：

[1] Novoselov K S, Geim A K, Morozov S V, *et al.* Electric field effect in atomically thin carbon films [J]. *Science.*, 2004, 306(5696): 666

[2] Castro Neto A H, Guinea F, Peres N M R, *et al.* The electronic properties of graphene [J]. Rev. Mod. Phys., 2009, 81(1): 109

[3] Fu Q, Bao X H. Progress in graphene chemistry [J]. *Chin. Sci. Bull.*, 2009, 54(18): 2657 (in Chinese) ,[傅强，包信和. 石墨烯的化学研究进展[J]. 科学通报, 2009, 54(18): 2657]

[4] Novoselov K, Jiang D, Schedin F, *et al.* Two-dimensional atomic crystals [J]. *Proc. Natl. Acad. Sci. U. S. A.*, 2005, 102(30): 10451

[5] Li D, Muller M B, Gilje S, *et al.* Processable aqueous dispersions of graphene nanosheets. *Nature Nano.*, 2008, 3(2): 101

[6] Berger C, Song Z, Li T, *et al.* Ultrathin epitaxial graphite: 2D electron gas properties and a route toward graphene-based nanoelectronics [J]. *J. Phys. Chem. B*, 2004, 108(52): 19912

[7] Cao H, Yu Q, Jauregui L A, *et al.* Electronic transport in chemical vapor deposited graphene synthesized on Cu: Quantum Hall effect and weak localization [J]. *Appl. Phys. Lett.*, 2010, 96(12): 122106

[8] Wintterlin J, Bocquet M L. Graphene on metal surfaces [J]. *Surf. Sci.*, 2009, 603(10-12): 1841

[9] Yu Q, Lian J, Siriponglert S, *et al.* Graphene segregated on Ni surfaces and transferred to insulators [J]. *Appl. Phys. Lett., 2*008, 93(11): 113103.

[10] Kim K S, Zhao Y, Jang H, *et al.* Large-scale pattern growth of graphene films for stretchable

transparent electrodes [J]. *Nature.*, 2009, 457(7230): 706

[11] Reina A, Thiele S, Jia X, *et al.* Growth of large-area single-and bi-layer graphene by controlled carbon precipitation on polycrystalline Ni surfaces [J]. *Nano Res.*, 2009, 2(6): 509

[12] Wang B, Bocquet M L, Marchini S, *et al.* Chemical origin of a graphene moiré overlayer on Ru (0001) [J]. *Phys. Chem. Chem. Phys.*, 2008, 10(24): 3530

[13] Xu Z, Buehler M J. Interface structure and mechanics between graphene and metal substrates: a first-principles study [J]. *J. Phys.: Condens. Matter*, 2010, 22(48): 485301.

[14] Zhang C J, Hu P. Methane transformation to carbon and hydrogen on Pd (100): pathways and energetics from density functional theory calculations [J]. *J. Chem. Phys.*, 2002, 116(1): 322

[15] Ciobîcă I M, Frechard F, Van Santen R A, *et al.* A DFT Study of Transition States for C-H Activation on the Ru (0001) Surface [J]. *J. Phys. Chem. B*, 2000, 104(14): 3364

[16] Watwe R M, Bengaard H S, Rostrup-Nielsen J R, *et al.* Theoretical studies of stability and reactivity of $CH_x$ species on Ni (111) [J]. *J. Catal.*, 2000, 189(1): 16

[17] Treier M, Pignedoli C A, Laino T, *et al.* Surface-assisted cyclodehydrogenation provides a synthetic route towards easily processable and chemically tailored nanographenes [J]. *Nature Chem.*, 2010, 3(1): 61

[18] Coraux J, N'Diaye A T, Busse C, *et al.* Structural coherency of graphene on Ir(111).[J]. *Nano Lett.*, 2008, 8(2): 565

[19] Loginova E, Bartelt N C, Feibelman P J *et al.* Evidence for graphene growth by C cluster attachment [J]. *New J. Phys.*, 2008, 10(9): 093026.

[20] Sprinkle M, Ruan M, Hu Y, *et al.* Scalable templated growth of graphene nanoribbons on SiC [J]. *Nature nano.*, 2010, 5(10): 727

[21] Günther S, Dänhardt S, Wang B, *et al.* Single Terrace Growth of Graphene on a Metal Surface [J]. *Nano Lett.*, 2011, 11(5): 1895

[22] Aradi B, Hourahine B, Frauenheim T. DFTB+, a sparse matrix-based implementation of the DFTB method [J]. *J. Phys. Chem. A*, 2007, 111(26): 5678

[23] Elstner M, Porezag D, Jungnickel G, *et al.* Self-consistent-charge density-functional tight-binding method for simulations of complex materials properties [J]. *Phys. Rev. B*, 1998, 58(11): 7260.

[24] Elstner M. The SCC-DFTB method and its application to biological systems [J]. *Theoretica Chimica Acta.*, 2006, 116(1): 316

[25] Yang M, Nurbawono A, Zhang C, *et al.* Two-dimensional graphene superlattice made with partial hydrogenation [J]. *Appl. Phys. Lett.*, 2010, 96(19): 193115.

[26] http://www.dftb.org/

[27] Hohenberg P, Kohn W. Inhomogeneous electron gas [J]. *Phys. Rev.*, 1964, 136(3B): B864

[28] Kong P, Wang H, Wang W, *et al.* First-principles calculations of electronic structure for B/N doped bilayer graphene [J]. *J. At. Mol. Phys.*, 2010, 27(4):775(in Chinese)[孔鹏，王菡，王薇，等．B/N掺杂对双层石墨烯电子特性影响的第一性原理研究[J].原子与分子物理学报, 2010, 27(4):775]

[29] Jiang F, He S. First-principles calculations of electronic properties for edge-modified graphene nanoribbons [J]. *J. At. Mol. Phys* 2011,28(6)(in Chinese)[江帆，何珊. 边缘修饰的石墨烯纳米带电子特性的第一性原理研究[J].原子与分子物理学报，2011，28(6)]

[30] Reich S, Maultzsch J, Thomsen C, *et al.* Tight-binding description of graphene [J]. *Phys. Rev. B*, 2002, 66(3): 035412

[31] Benaand C, Montambaux G. Remarks on the tight-binding model of graphene [J]. *New J. Phys.*, 2009, 11(9): 095003

[32] Hu H X, Zhang Z H, Liu X H, *et al.* Tight binding studies on the electronic structure of graphene nanoribbons [J]. *Acta. Phys. Sin.*, 2009, 58(10): 7156(in Chinese)[胡海鑫，张振华, 刘新海，等. 石墨烯纳米带电子结构的紧束缚法研究[J].物理学报，2009，58（10）：7156]

[33] Bertoni G, Calmels L, Altibelli A, *et al.* First-principles calculation of the electronic structure and EELS spectra at the graphene/Ni(111) interface [J]. *Phys. Rev. B*, 2005, **7**1(7): 075402.